\documentclass[letterpaper, 10 pt, conference]{ieeeconf}
\IEEEoverridecommandlockouts
\usepackage{times}
\usepackage{fancyhdr}
\usepackage{float}
\usepackage{cite}
\usepackage{listings}
\usepackage{booktabs}
\usepackage{subfig, bm}

\usepackage{amsthm}
\usepackage{tikz}
\usepackage{amssymb}
\usepackage{mdwlist}
\usepackage{hyperref}
\hypersetup{colorlinks,linkcolor={blue},citecolor={blue}} 
\usepackage{amsmath,graphicx,xcolor}
\usepackage{subfiles}
\usepackage{import}
\usepackage{dsfont}
\usepackage{xfrac}
\usepackage[font=small,labelfont=bf]{caption}
\usepackage{mathtools}
\usepackage{algorithm}
\usepackage{algorithmic}
\usepackage{adjustbox}
\usepackage{pifont}
\usepackage{sjmacros}
\usepackage{thmstyles}
\usepackage{stfloats}
\usepackage{cuted}

\title{\LARGE \bf
A Stochastic Robust Adaptive Systems Level Approach to Stabilizing Large-Scale Uncertain Markovian Jump Linear Systems
}

\author{
SooJean Han$^{1\dagger}$, Minwoo M. Kim$^{1}$, Ieun Choo${^2}$
\thanks{Korea Advanced Institute of Science \& Technology (KAIST), S. Korea. $^{1}$School of Electrical Engineering. $^{2}$Department of Aerospace Engineering.}
\thanks{$^{*}$Emails: \{\texttt{soojean}, \texttt{epsilon4b}, \texttt{beckstar1}\}\texttt{@kaist.ac.kr}.}}

\begin{document}
\maketitle
\thispagestyle{plain}
\pagestyle{plain}
\allowdisplaybreaks

\begin{abstract}
    We propose a unified framework for robustly and adaptively stabilizing large-scale networked uncertain Markovian jump linear systems (MJLS) under external disturbances and mode switches that can change the network's topology.
Adaptation is achieved by using minimal information on the disturbance to identify modes that are consistent with observable data.
Robust control is achieved by extending the system level synthesis (SLS) approach, which allows us to pose the problem of simultaneously stabilizing multiple plants as a two-step convex optimization procedure.
Our control pipeline computes a likelihood distribution of the system's current mode, uses them as probabilistic weights during simultaneous stabilization, then updates the likelihood via Bayesian inference.
Because of this ``softer’’ probabilistic approach to robust stabilization, our control pipeline does not suffer from abrupt destabilization issues due to changes in the system's true mode, which were observed in a previous method.
Separability of SLS also lets us compute localized robust controllers for each subsystem, allowing for network scalability; we use several information consensus methods so that mode estimation can also be done locally.
We apply our algorithms to disturbance-rejection on two sample dynamic power grid networks, a small-scale system with 7 nodes and a large-scale grid of 25 nodes.


\end{abstract}
\vskip.1cm


\section{Introduction}
Markovian jump systems are often used to represent systems driven by a process of jumps which define a type of switching among multiple modes.
Many applications in various branches of engineering can be modeled using Markovian jump systems, especially fault-tolerant control systems (in which jumps arise due to sudden faults) or a system component which evolves in discrete intervals (e.g., a thermostat which switches between ``heating'' and ``cooling'').
For simplicity of analysis, these jumps are usually modeled with a Markov chain, although this assumption turns out to work well in practice since the switching points are usually nearly memoryless (e.g., the current switching time does not depend on the past switching times).

An especially important class of real-world systems that can be modeled using Markovian jump systems are large-scale networks with dynamic topologies, such as formation-flying UAVs, electric power grids, or transportation networks.
Although much of the past literature on Markovian jump linear system (MJLS) control is implemented in a centralized fashion (e.g.,~\cite{aberkane2011,matei2008,deSouza2006}), a distributed approach may be more suitable for large-scale networks, which requires each subsystem to rely only on local information.
Reliance on local information makes it more probable that many of the system's parameters (e.g., the current mode) will be uncertain or unknown; this can make stabilization difficult without using system identification.
Moreover, being able to stabilize the system without full identification requires the controller to be robust, but this is difficult when the topology (i.e., the way control signals propagate throughout the physical network) is unknown.

Motivated by the concerns above, this paper proposes a unified framework for robustly and adaptively stabilizing large-scale MJLS under external disturbances and unknown mode changes. 
Though most literature focuses on parametric changes, we address the case where the mode switches correspond to changes in the network's topology (i.e., the support of the dynamics matrix $A$).
Adaptation is achieved by using minimal knowledge about the external disturbance to estimate modes that are consistent with the system data (i.e., state and control trajectories).
In mode estimation, we assign probabilistic weights to each mode based on how likely it is equal to the system's true mode, then updates the likelihood via Bayesian inference for the next iteration.
Robust control is achieved by extending system level synthesis (SLS)~\cite{anderson19} to use the likelihood distribution computed by mode estimation as priority weights in the simultaneous stabilization problem.

There are several key benefits of our approach.
Compared to book-keeping a set of consistent modes (see~\cite{han20l4dc}), the Bayesian mode estimation procedure can be viewed as a softer approach to eliminating inconsistent modes and allows us to remove any strict norm-boundedness assumptions on the external disturbance; we show here that the two methods can be combined to further reduce uncertainty in the true mode.
Due to the ability of SLS to include constraints, simultaneous stabilization of multiple plants can be posed as a two-step convex optimization problem which is easily solvable using numerical algorithms (e.g., CVX).
This approach to robust stabilization allows our controller to stabilize the system for all time (not experiencing sudden destabilization whenever the system's ground-truth mode changes).
We demonstrate these effects by simulating various versions of our framework to the concrete application of disturbance-rejection and fault-tolerance in dynamic power grid networks.
Separability of SLS also lets us compute localized robust controllers for each subsystem, and we introduce several information consensus methods so that mode estimation can also be localized, allowing for scalability of the overall pipeline.

Our paper is organized as follows.
The problem is set up in~\sec{setup}.
A brief review of all the background and relevant preliminaries (MJLS stability, SLS, and consensus) is done in~\sec{background}.
Our main control framework is introduced in~\sec{method}, with mode estimation in Section~\ref{subsec:csn} and the stochastic multimode stabilization in Section~\ref{subsec:multimode_sls}.
Numerical simulations are done in~\sec{simulation}.
We conclude the paper in~\sec{conclusion}.


\section{Problem Setup}\label{sec:setup}
We are concerned with MJLS of the following form:
\begin{align}\label{eq:plant_dynamics}
    \hskip1cm\xvect[t+1] &= A(\xi_t)\xvect[t] + B(\xi_t)\uvect[t] + \wvect[t]
\end{align}
Here, $\xvect[t]{\,\in\,} \Rbb^{n_x}$ is the state, $A(\xi_t){\,\in\,}\Rbb^{n_x\times n_x}$ is the dynamics matrix which changes according to the mode variable $\xi_t$, and $\uvect[t]{\,\in\,}\Rbb^{n_u}$ is the control input.
The external noise process $\wvect[t]{\,\in\,}\Rbb^{n_x}$ is unknown at each time $t$.
The mode process $\{\xi_t\}_{t=1}^{\infty}$ takes values from the set $\Xcal{\,\triangleq\,}\{1, \cdots, M\}$, where $M{\,\in\,}\Nbb$, and is defined such that $\xi_t{\,:\,}\Omega{\,\to\,}\Xcal$ on probability space $(\Omega, \Fcal, \Pbb)$ with filtration $\{\Fcal_t\}_{t=0}^{\infty}$, $\Fcal_t{\,\triangleq\,}\sigma(\xi_0, \xi_1, \cdots, \xi_t)$.
Moreover, $\{\xi_t\}_{t=0}^\infty$ is generated from a Markov chain over $\Xcal$ with transition probability matrix (TPM) denoted by $P{\,\in\,}\Rbb^{M\times M}$.

We focus primarily on the case where the mode switches represent topological changes in the network (e.g., line disconnections in power grids, road obstructions in transportation networks). 
Parametric uncertainties can be viewed as a special case which do not cause major changes to the supports of the system matrices ($A$ or $B$).
At each time $t$, we assume the current state $\xvect[t]$ can be observed and the current mode $\xi_t$ is unknown.
We assume for simplicity that the different possible modes $\Xcal$ are known; unknown or time-varying $\Xcal$ is a subject of future work.

We focus on the case where the dynamics~\eqn{plant_dynamics} is given by a network dynamics on $M$ possible directed graphs $\Gcal(m){\,=\,}(\Vcal,\Ecal(m))$ without loops and shared vertex set $\Vcal$:
\begin{align}\label{eq:plant_dynamics_local}
    \xvect_i[t+1] &= A_{(ii)}(\xi_t)\xvect_i[t] \\ 
    &+ \sum\limits_{j\in\Ncal_i(\xi_t)}A_{(ij)}(\xi_t)\xvect_j[t] + B_{(ii)}(\xi_t)\uvect_i[t] + \wvect_i[t]\notag
\end{align}
Here, $\xvect_i[t]{\,\in\,}\Rbb^{n}$ is the local state, $\abs{\Vcal}{\,\triangleq\,}N$ is the number of nodes (i.e., $n_x{\,=\,}Nn$), and $\Ncal_i(m){\,\subseteq\,}\Vcal$ is the set of inward neighbors of $i{\,\in\,}\Vcal$ under mode $m$.
The $A_{(ii)}, A_{(ij)}$ and $B_{(ii)}$ are the corresponding submatrices of $A$ and $B$ respectively; in particular, the $B$ matrix is \emph{block diagonal}.
Because we focus on localized controller implementations for large-scale systems, each subsystem $i$ has its own local control input $\uvect_i[t]\in\Rbb^m$ (i.e., $n_u=Nm$).


\section{Background}\label{sec:background}
\subsection{Preliminaries on MJLS Stability}\label{subsec:mjls}
%
MJLS~\eqn{plant_dynamics} 
is \textit{stochastically stable} if $\Ebb[\sum_{t=0}^{\infty} \norm{\xvect[t]}^2]{\,<\,}\infty$
for any initial conditions $\xvect_0{\,\in\,}\Rbb^{n_x}, \xi_0{\,\in\,}\Xcal$~\cite{costa_book}.
%
When the scale of the system is small enough for a centralized controller implementation, one common way of verifying stochastic stability (when $\uvect[t]{\,=\,}0$) is to check the existence of positive-definite $\{S_m, m\in\Xcal\}$ such that Lyapunov-like inequalities are satisfied: $A_m^{\top}\sum_{j\in\Xcal}P_{mj}S_j A_m - S_m {\,\prec\,}0$ for all $m{\,\in\,}\Xcal$~\cite{costa_book}.
Many feedback control approaches for MJLS in the literature compute a time-varying, mode-dependent state-feedback gain $K_{t,m}$ such that $\uvect[t]{\,=\,}K_{t,\xi_{t-1}}\xvect[t]$; ref.~\cite{matei2008}, for example, achieves this via a dynamic programming approach
to solve the LQR control problem for each mode.
However, because these approaches rely on a centralized controller implementation, they do not scale to large-dimensional systems with a large number of modes.
Moreover, these methods typically operate under the assumption that the current mode $\xi_t$ is known at each time $t$.

\subsection{System Level Synthesis}\label{subsec:sls}
System level synthesis (SLS)~\cite{wang19,anderson19} is a framework for the controller synthesis of linear, discrete time network systems with static topologies $\Gcal{\,\triangleq\,}(\Vcal, \Ecal)$, expressed as
\begin{align}\label{eq:linear_dt_static}
    \xvect[t+1] = A\xvect[t] + B\uvect[t] + \wvect[t]
\end{align}
Optimal control problems are associated with~\eqn{linear_dt_static}, which can be posed in the general form
\begin{align}\label{eq:gen_opt_ctrl}
    \min_{\uvect[\cdot]} f(\xvect, \uvect) \text{ subject to }\eqn{linear_dt_static}.
\end{align}
Common examples of~\eqn{gen_opt_ctrl} are prevalent in the literature depending on the properties of $\wvect[t]$.
Let $h(\xvect,\uvect)\triangleq \sum_{t=0}^T\xvect^{\top}[t]Q\xvect[t] + \uvect^{\top}R\uvect[t]$ for some horizon $T{\,>\,}0$ and matrices $Q,R{\,\succ\,}0$.
LQG $(\Hcal_2)$ control is posed as $f(\xvect, \uvect){\,=\,}\Ebb[h(\xvect, \uvect)]$ when $\wvect[t]{\,\sim\,}\Ncal(0,\Sigma_w)$, $\Hcal_{\infty}$ control is when $f(\xvect, \uvect){\,=\,}\max_{\norm{\wvect[t]}_2\leq 1}h(\xvect, \uvect)$ and $\norm{\wvect[t]}_2{\,\leq\,}1$, and $\Lcal_1$ control is when $f(\xvect, \uvect){\,=\,}\max_{\norm{\wvect[t]}_{\infty}\leq 1}h(\xvect, \uvect)$ and $\norm{\wvect[t]}_{\infty}{\,\leq\,}1$ for all $t$; see~\cite{anderson19}.

For linear state-feedback optimal control problems, laws of the form $\uvect[t]{\,=\,}K\xvect[t]$ are synthesized via a gain $K$ which are computed via common matrix algorithms (e.g., Riccati equation, LMI characterization).
A key component in the SLS approach is that instead of synthesizing the gain $K$, it synthesizes the entire closed-loop response via $\boldsymbol{\Phi}{\,\triangleq\,}\{\boldsymbol{\Phi}_x,\boldsymbol{\Phi}_u\}$ such that $\xvect{\,=\,}\boldsymbol{\Phi}_x\wvect$ and $\uvect{\,=\,}\boldsymbol{\Phi}_u\wvect$.
This parametrization provides important advantages such as scalability to large networks and the ability to include more controller constraints in a convex manner (i.e., performance requirements, sparsity structure).
We will only review the parts of SLS needed for the construction of our approach, but for more details, we refer the reader to~\cite{wang19,anderson19}.

\begin{lemma}[Thm. 4.1 in~\cite{anderson19}]
    For static network dynamics~\eqn{linear_dt_static}, the following are true.
    First, the affine subspace described by 
    \begin{align}\label{eq:achievable_sls}
        \begin{bmatrix}
            zI - A & -B
        \end{bmatrix}\begin{bmatrix}
            \boldsymbol{\Phi}_x\\ \boldsymbol{\Phi}_u
        \end{bmatrix} = I, \hskip.2cm \boldsymbol{\Phi}_x, \boldsymbol{\Phi}_u\in\frac{1}{z}\Rcal\Hcal_{\infty}
    \end{align}
    parametrizes all possible system responses $\boldsymbol{\Phi}$ achievable by an internally stabilizing state feedback controller, where $(1/z)\Rcal\Hcal_{\infty}$ is the set of stable, strictly proper rational transfer matrices. Conversely, for any $\boldsymbol{\Phi}$ which satisfies the condition in~\eqn{achievable_sls}, $K{\,\triangleq\,}\boldsymbol{\Phi}_u\boldsymbol{\Phi}_x^{-1}$ achieves the desired system responses $\boldsymbol{\Phi}_x$ and $\boldsymbol{\Phi}_u$, possibly by an internally stabilizing implementation.
\end{lemma}

The state-feedback control law is then implemented as:
\begin{align}\label{eq:sf_implement}
    \hat{\xvect}[t] &= \sum\limits_{s=2}^T \Phi_x[s] \boldsymbol{\delta}[t+1-s],\hskip.1cm \boldsymbol{\delta}[t] = \xvect[t] - \hat{\xvect}[t]\\
    \uvect[t] &= \sum\limits_{s=1}^T \Phi_u[s] \boldsymbol{\delta}[t+1-s]\notag
\end{align}
where $\boldsymbol{\Phi}_x = \sum_{\tau=1}^{T} \Phi_x[\tau]z^{-\tau}$ and similarly for $\boldsymbol{\Phi}_u$, $\boldsymbol{\delta}$ is the controller's \textit{internal state}, $T{\,\in\,}\Nbb$ is a user-chosen \emph{finite} horizon which further restricts the controller.

The generic optimal control problem~\eqn{gen_opt_ctrl} can now be posed using the SLS framework~\cite{anderson19}:
\begin{align}\label{eq:gen_opt_ctrl_sls}
    \min_{\boldsymbol{\Phi}} g(\boldsymbol{\Phi}) \text{ s.t. }\eqn{achievable_sls}, \Phi_x[\tau] = \mathbf{0}, \Phi_u [\tau]=\mathbf{0} \text{ for } \tau >T, \hskip.1cm \boldsymbol{\Phi}{\,\in\,}\Scal
\end{align}
where $\Scal$ is a set of additional convex constraints.

To obtain a distributed implementation of~\eqn{sf_implement}, an important property to consider is \textit{separability} of the centralized problem~\eqn{gen_opt_ctrl_sls}.
Columnwise separability~\cite{wang18} with respect to some columnwise partition $\{c_1, \cdots, c_N\}$ means 1) we can rewrite 
\begin{equation} \label{eq:separable_costs}
    g(\boldsymbol{\Phi}){\,=\,}\sum_j g_j(\boldsymbol{\Phi}(:,c_j))
\end{equation}
for some functionals $g_j$ related to $g$, and 2) there exists some partition $\Scal_1\cup\cdots\cup\Scal_N$ of the constraint set $\Scal$ such that $\boldsymbol{\Phi}{\,\in\,}\Scal$ iff $\boldsymbol{\Phi}(:,c_j){\,\in\,}\Scal_j$ for all $j$.
Some examples of columnwise-separable functions $g$ are the squared Frobenius norm or its weighted variations, and examples of columnwise-separable constraints $\Scal$ are matrix support constraints.
In our notations, we add a subsystem index $i$ such that $\boldsymbol{\Phi}$ is implemented as $\boldsymbol{\Phi}_i{\,\triangleq\,}\{\Phi_{x,i}[s], \Phi_{u,i}[s]\}_{s=1}^T$ for each node $i{\,\in\,}\Vcal$.
%

\subsection{Information Consensus}\label{subsec:consensus_bg}
In dynamic topology networks, the SLS-based controller described in Section~\ref{subsec:sls} cannot be implemented without full knowledge of the true current mode $\xi_t$.
While each subsystem can estimate the mode exclusively from its own local data, it may be less accurate than combining the local data with additional data from its neighbors.
This motivates the inclusion of \textit{information consensus}~\cite{olfati04,xiao05} in our approach, which, if the communication graph changes over time, is possible under conditions such as joint-connectedness.
For simplicity throughout this paper, we will focus on networks with a static, connected undirected communication graph.


\section{Main Controller Framework}\label{sec:method}
Our general framework consists of two main phases.
In the first phase, \textit{mode estimation}, each subsystem constructs an estimate of the true mode using local data and information consensus with its neighbors.
The second phase, \textit{stochastic multimode SLS}, is used to control the system before the true mode is fully identified; we design a robust controller which can be used to simultaneously stabilize multiple modes in some uncertainty set returned by the mode estimation phase.
Various specific versions of the general pipeline are organized in~\tab{compare_architectures}.

\subsection{Mode Estimation}\label{subsec:csn}
In general, the performance of most robust controllers is largely dependent on the size of the uncertainty set (i.e., how well we understand the uncertainties of the system).
We include an adaptive, time-varying mode estimation process, where system data (i.e., state and control trajectories) is gathered to reduce the uncertainty surrounding the ground-truth mode.
We consider two types of robustness measures: 1) against the \textit{most likely} disturbance, when the distribution of $\wvect[t]$ is known (e.g., Gaussian), and 2) against the \textit{worst-case} disturbance, when only the norm-bound of $\wvect[t]$ is known.

\bgroup
\def\arraystretch{1.2}
\begin{table}
    \begin{adjustbox}{width=0.95\columnwidth,center}
      \begin{tabular}{| c | c | c | c | c |}
        \hline
        \textbf{No.} &\textbf{Multimode SLS} & \textbf{Consistency} & \textbf{Consensus} \\ \hline \hline
         1 & StochAdapt & Infnorm CSN & Centralized \\ \hline
         2 & StochAdapt & Infnorm CSN & Decentral-Separate \\ \hline
         3 & StochAdapt & Infnorm CSN & Decentral-Merged \\ \hline
         4 & StochAdapt & Bayesian & Decentral-Separate\\ \hline
         5 & FullRobust & -- & -- \\ \hline
      \end{tabular}
    \end{adjustbox}
    \caption{The different control architectures considered in the paper.
    No. 2-4 are our main methods.
    No. 1 extends~\cite{han20l4dc} and No. 5 comes from Rmk.~\ref{rmk:fullrobust}; they are used for comparisons in Sec.~\ref{sec:simulation}.}
    \vspace{-.3cm}
    \label{tab:compare_architectures}
\end{table}
\egroup

\noindent\textbf{Consistent Set Narrowing} (\texttt{Infnorm CSN} in~\tab{compare_architectures}):\\
Consistent set narrowing (CSN) is used to accommodate worst-case disturbance, when we can assume there exists a $\overline{w}{\,>\,}0$ such that $\wvect[t]$ satisfies the bound $\norm{\wvect}_{\infty}{\,\leq\,}\overline{w}$ for all $t$.
Because we cannot observe $\xi_t$ directly, we perform an identification step using state and input trajectories as data.
At each timestep, we keep track of a set of modes $\Ccal[t]$ that are consistent with $(\xvect[t-1],\uvect[t-1],\xvect[t])$; $\Ccal[t]$ is hereby called the \textit{consistent set}.
%
The dynamics~\eqn{plant_dynamics} can be manipulated into a condition:
\begin{align}\label{eq:infnorm}
    \norm{\xvect[t] - A(m)\xvect[t-1] - B(m)\uvect[t-1]}_{\infty} \leq \overline{w}
\end{align}
and the consistent set is iteratively defined as
\begin{align}\label{eq:infnorm_iter}
    &\Ccal[t] \triangleq\\
    &\hskip.2cm\begin{cases}
        \left\{m\in\Ccal[t-1] \hskip.1cm\middle|\hskip.1cm \eqn{infnorm} \text{ is satisfied}\right\} &\text{ if it is nonempty}\\
        \Xcal &\text{ else,}
    \end{cases}\notag
\end{align}
with $\Ccal[0]{\,=\,}\{\xi_0\}$ (assumed known initial mode). 
That is, we reset $\Ccal[t]$ to $\Xcal$ if all modes from $\Ccal[t-1]$ are inconsistent, as it would imply that a mode transition has occurred.
Note that when the mode process $\{\xi_t\}$ does not follow a Markov chain, we can use a different estimation method based on learning recurrent patterns in $\{\xi_t\}$, inspired by~\cite{han23auto}.
This is a subject of future work.

\noindent\textbf{Bayesian Consistency} (\texttt{Bayesian} in~\tab{compare_architectures}):\\
We propose a novel alternative measure of consistent modes, Bayesian consistency, which assigns probabilistic weights based on the most likely mode the system is currently in.
The interpretation of these weights is that the robust controller assigns higher priority to develop a stabilizing law for modes with larger weights.
Bayesian consistency can be implemented when the distribution of $\wvect[t]$ is known, which lets us generalize to noise profiles that are not norm-bounded.

Each subsystem $i$ at time $t$ assigns likelihood probabilities $\pvect_i[t]{\,\triangleq\,}(p_{i,1}[t],\cdots,p_{i,M}[t])^{\top}$, where $p_{i,m}[t]$ is the inferred probability of the system being in mode $m{\,\in\,}\Xcal$ at time $t$.
For each $i$, $\pvect_i[t{\,+\,}1]$ is updated from $\pvect_i[t]$ via a form of Bayesian inference.
In the concrete case where the noise injected into the $i$th subsystem $\wvect_i[t]{\,\sim\,}\Ncal(0,\Sigma_w)$ is mean-zero Gaussian with $\Sigma_w{\,\succ\,}0$, a normalized estimate of the current noise is computed using the dynamics equation:
\begin{align}\label{eq:pval_w}
    \tilde{\wvect}_{i,m}[t] &\approx \Sigma_w^{-1/2}*\\
    &\big(\xvect_i[t] - A_{(i \cdot)}(m)\xvect[t-1] - B_{(i \cdot)}(m)\uvect[t-1]\big),\notag
\end{align}
where $A_{(i \cdot)}{\,\in\,}\Rbb^{n\times n_x}$ is the submatrix of $A$ formed by taking the rows affiliated with subsystem $i$, and likewise for $B_{(i \cdot)}$.
For each $m$, p-value $\texttt{p}_{i,m}$ is computed using $\tilde{\wvect}_{i,m}[t]$; then, if $\sum_m p_{i,m}[t] \texttt{p}_{i,m}$ is less than a chosen threshold $\varepsilon$ and the likelihood probabilities $p_i[t]$ have not been ``reset'' for some number $R$ of timesteps, then we ``reset'' the likelihood probabilities to be uniformly-distributed for time $t+1$.
The ``resetting'' mechanism is implemented for reasons similar to particle deprivation in sequential importance resampling~\cite{doucet2008}, where there may be no particles nearby the correct mode.
Otherwise, we recompute the probabilities as
\begin{align}\label{eq:bayes_update}
    p_{i,m}[t+1] &= p_{i,m}[t]\left(\epsilon + e^{-\frac{1}{2}\tilde{\wvect}_{i,m}^\top[t]\tilde{\wvect}_{i,m}[t]}\right)\hskip.2cm\forall\hskip.1cm m\in\Xcal
\end{align}
where $\epsilon$ is chosen small for numerical stability and we normalize $p_{i,m}[t+1]{\,\to\,}p_{i,m}[t+1]/(\sum_mp_{i,m}[t+1])$ after computing~\eqn{bayes_update}.
Note that this method also applies to non-Gaussian noise distributions as long as their probability density functions can be reasonably estimated a priori.

\subsection{Incorporating Consensus}\label{subsec:consensus}
In order to take full advantage of a distributed controller implementation (to be described in Section~\ref{subsec:multimode_sls}), each subsystem must estimate the mode using only local information.
Each subsystem $i\in\Vcal$ keeps track of its own consistency measure (i.e., local consistent set $\Ccal_i[t]$ in \texttt{Infnorm CSN}, or local probability vector estimate $\pvect_{i}[t]$ in \texttt{Bayesian}).

We consider two main cases of implementing consensus.
First, we use \textit{centralized consensus} (\texttt{Centralized} in~\tab{compare_architectures}), in which all subsystems simply maintain a single global consistency measure (e.g., $\Ccal_i[t]{\,\equiv\,}\Ccal[t]$ for all $i$); this is described exactly by~\eqn{infnorm} and~\eqn{infnorm_iter}.
Second, we take a \textit{decentralized consensus} approach, in which there are two further subcases to consider.
In the first subcase (\texttt{Decentral-Separate} in Table~\ref{tab:compare_architectures}), each subsystem keeps track of its own consistency measure and updates it over time via~\eqn{infnorm_iter}.
For \texttt{Infnorm CSN}, this is done with $\Ccal_i[t]$, updated via
\begin{align}\label{eq:infnorm_local}
    \norm{\xvect_{i}[t] - A_{(i\cdot)}(m)\xvect[t-1] - B_{(i\cdot)}(m)\uvect[t-1]}_{\infty} \leq \overline{w}
\end{align}
in place of~\eqn{infnorm} (i.e., it is a localized version of the condition).
For \texttt{Bayesian}, this is done exactly as described in~\eqn{bayes_update}.
In the second subcase (\texttt{Decentral-Merged} in Table~\ref{tab:compare_architectures}), for Infnorm CSN, subsystems exchange consistency measures with their neighbors in an undirected communication graph, which we introduce as $\Gcal^c{\,\triangleq\,}(\Vcal,\Ecal^c)$.
For simplicity, we assume the topology of this communication network is unchanging and different from the physical network $\Gcal(m)$, i.e., even if two nodes $i,j{\,\in\,}\Vcal(m)$ are physically disconnected ($(i,j){\,\not\in\,}\Ecal(m)$), they may still be able to communicate with each other ($(i,j){\,\in\,}\Ecal^c$).
Each subsystem then keeps track of a \textit{merged} consistent set which is used to construct the local control signal:
\begin{align}\label{eq:merged_consistency_infnorm}
    \overline{\Ccal}_i[t] \triangleq \begin{cases}
        \bigcap_{j\in\Ncal^c_i \cup \{i\}}\Ccal_j[t] &\text{ if } \bigcap_{j\in\Ncal^c_i \cup \{i\}} \Ccal_j[t]\neq\varnothing\\
        \Xcal &\text{ else}
    \end{cases}
\end{align}
Here, $\Ncal^c_i$ is the set of all neighbors of node $i$ in graph $\Gcal^c$.

Consensus is also used to develop an aggregated estimate $\hat{\xi}_t$ of the true mode $\xi_t$.
For the Infnorm case, $\hat{\xi}_t$ is simply the most frequent mode in the set $\cup_i\Ccal_i[t]$ (or $\cup_i\overline{\Ccal}_i[t]$ in the merged consensus subcase).
For the Bayesian case, $\hat{\xi}_t$ is the most frequent mode among $\hat{\xi}_{i,t}$ for $i{\,\in\,}\Vcal$, where $\hat{\xi}_{i,t}{\,\triangleq\,}\operatorname{argmax}_m p_{i,m}[t]$.
While $\hat{\xi}_t$ is a good metric of performance (and used in Section~\ref{sec:simulation}), it is not actually used to compute the control law at each time, which incorporates the consistent sets $\Ccal_i[t]$ and/or probabilities $\pvect_i[t]$ directly; we will see this in the next subsection.

\begin{remark}[Consistency Measure Relationship]\label{rmk:consistency}
    We can represent each consistent set $\Ccal_i[t]$ in terms of the likelihood probabilities $\pvect_i[t]$ via:
    \begin{align*}
        p_{i,m}[t] = \begin{cases}
            1/\abs{\Ccal_i[t]} &\text{ if } m\in\Ccal_i[t]\\
            0 &\text{ else}
        \end{cases}
    \end{align*}
    and likewise for $\overline{\Ccal}_i[t]$ if \texttt{Decentral-Merged} is being used.
    This representation will be key in our control law design, where the same two-step optimization problem can be used despite the different consistency measures.  
    Moreover, compared to \texttt{Infnorm CSN}, \texttt{Bayesian} is a softer process of eliminating inconsistent modes, keeping $\Ccal_i[t]{\,=\,}\Xcal$ for all nodes $i$ and time $t$, and instead using probabilistic weights to assign priority to the most consistent modes. 
    As we see in~\sec{simulation}, integrating these priority weights into our control pipeline enables us to soften the mode estimation phase and stabilize the system without experiencing abrupt instabilities whenever the mode changes in the system, a phenomenon that was observed previously in~\cite{han20l4dc}.
\end{remark}

\subsection{Stochastic Multimode SLS}\label{subsec:multimode_sls}
Simultaneous stabilization of multiple discrete-time LTI systems has been studied extensively in the past.
There are two main types of methods: 1) representing each plant via the Youla-Kucera parametrization and designing a single compensator such that there are no pole-zero cancellations (e.g.,~\cite{blondel92}), and 2) iteratively solving  collection of LMI or Riccati-like equations until we converge towards a single feedback gain (e.g.,~\cite{cao99}).
It is generally known that simultaneous stabilization of three or more plants is difficult using the first type of methods~\cite{blondel94_book,fonte2001}.
Moreover, it is not immediately obvious how either type of method can be extended to a distributed implementation.

We bypass the issues mentioned above by posing simultaneous stabilization as a convex optimization problem to construct a single control law for the entire MJLS.
Our method extends and reformulates the original SLS problem~\eqn{gen_opt_ctrl_sls} for MJLS~\eqn{plant_dynamics}, with dynamic topologies $\Gcal(m){\,=\,}(\Vcal,\Ecal(m))$, $m{\,\in\,}\Xcal$.
Combined with scalability to larger systems, our method makes the simultaneous stabilization problem more tractable to solve compared to prior approaches.

One key distinction between standard SLS (Section~\ref{subsec:sls}) and our proposed multimode algorithms (\texttt{StochAdapt} in~\tab{compare_architectures}) are the support constraints, especially since the topology of the system could change according to the mode process.
Prior work~\cite{han20l4dc} had proposed to change the support constraints based on each subsystem's current estimate of the true mode, but this requires a good estimate $\hat{\xi}_t$ of the true current mode.
Our approach here is to keep the supports constant throughout time by aggregating the possible topologies of the system.
\begin{align}\label{eq:support_constraints}
    &S\triangleq\bigcup_{s=1}^t \frac{1}{z^s}\left(\bigcup_{\ell=0}^h\operatorname{adj}\left(\Gcal^c\right)^\ell\cap\left(\bigcup_{\ell=0}^h\bigcup_{m\in\Xcal} \operatorname{adj}\left(\Gcal(m)\right)^\ell\right)\right)\\
    &\operatorname{supp}\left( \Phi_x[t]\right) = S\otimes \mathbf{1}^{n\times n},\quad \operatorname{supp}\left( \Phi_u[t]\right) = S\otimes \mathbf{1}^{n\times m}\notag
\end{align}
where $\operatorname{supp}(\cdot)$ denotes the support operator, $\operatorname{adj}$ denotes the adjacency matrix, $\mathbf{1}^{k\times \ell}$ is the $k{\,\times\,}\ell$ matrix of all ones, $\otimes$ is the Kronecker product, and $h{\,\in\,}\Nbb$ is the number of hops allowed for communication.

Prior MJS literature~\cite{costa95,costa98} may lead one to simply combine the achievability constraints~\eqn{achievable_sls} over all possible modes ($[zI - A(m), -B(m)][\boldsymbol{\Phi}_x^{\top}, \boldsymbol{\Phi}_u^{\top}]^{\top}{\,=\,}I$ for all $m{\,\in\,}\Xcal$), but these constraints often cannot be satisfied with perfect equality using a single $\boldsymbol{\Phi}$.
This motivates our design of a \textit{robustness margin}, which is minimized by performing a \emph{two-step} optimization to solve for the optimal $\boldsymbol{\Phi}_i^{(t)}{\,=\,}\{ \Phi_{x,i}^{(t)} [\tau], \Phi_{u,i}^{(t)}[\tau]\}_{\tau=1}^T$ for each subsystem $i$ at time $t$.
For scalability, our optimization problem is posed columnwise-separable (see~\eqn{separable_costs}) so that $\Phi_{x,i}^{(t)} [\tau]{\,\in\,}\Rbb^{n_x\times n}$ and $\Phi_{u,i}^{(t)} [\tau]{\,\in\,}\Rbb^{n_u\times n}$ for each $i{\,\in\,}\Vcal$ and $\tau{\,=\,}1,\cdots, T$.
The first step computes a vector of \textit{robustness margins} $\lambda_i[t]{\,\triangleq\,}[\lambda_{i,1}[t], \cdots,\lambda_{i,M}[t]]^{\top}$ by minimizing $\sum_{m=1}^M p_{i,m}[t] \lambda_{i,m}[t]$ for which the feasibility constraints are satisfied for all modes:
\begin{align}\label{eq:robust_constraint}
    \sum\limits_{s=1}^T \norm{\Delta_{i,s}(A(m), B(m), \Phi_x^{(t)}, \Phi_u^{(t)})} \leq \lambda_{i,m}[t], \hskip.1cm\forall m\in\Xcal
\end{align}
where $\Delta_{i,s}(A, B, \Phi_x, \Phi_u)\triangleq(\Phi_{x,i}[s{\,+\,}1] - A\Phi_{x,i}[s] - B\Phi_{u,i}[s])$ and we set $\Phi_x[T+1]=\mathbf{0}$.

The second step uses the optimal value $\lambda^*$ from step 1, and minimizes a performance cost $g(\cdot)$ in terms of $\boldsymbol{\Phi}^{(t)}=\{\Phi_x^{(t)} [\tau], \Phi_u^{(t)} [\tau]\}_{\tau=1}^T$.
Together, the procedure becomes:
\begin{subequations}\label{eq:StochAdapt}
    \begin{align}
        &\min_{\lambda_{i}[t], \boldsymbol{\Phi}_i^{(t)}} f(\lambda_{i}[t], \boldsymbol{\Phi}_i^{(t)})\\
        \text{ s.t. } &~\eqn{support_constraints},~\eqn{robust_constraint},\hskip.1cm\Phi_{x,i}^{(t)}[1] = I_{(\cdot i)}\\
        &\left(\sum_{m\in\Xcal} p_{i,m}[t]\lambda_{i,m}[t]\leq 2 \lambda^*\right)\cdot\mathds{1}\{\text{step 2}\}\label{eq:robust_margin_bound}\\
        &\norm{\Phi_{x,i}^{(t-1)}[s] - \Phi_{x,i}^{(t)}[s]} \leq \rho^*\hskip.2cm \forall\hskip.1cm s=1,\cdots,T\label{eq:adaptation_constraint} 
    \end{align}
\end{subequations}
where $I_{(\cdot i)}$ are the columns of the identity matrix corresponding to node $i$, $\rho^* >0$ is also user-chosen, and
\begin{align}\label{eq:stochadapt_obj}
    f(\lambda_i [t], \boldsymbol{\Phi}_i^{(t)}) = \begin{cases}
        \sum_{m\in\Xcal} p_m[t]\lambda_m[t] &\text{ if step 1}\\
        g_i(\boldsymbol{\Phi}_i^{(t)}) \text{ from}~\eqref{eq:separable_costs} &\text{ if step 2}
    \end{cases}
\end{align}
Here, $\pvect_{i}[t]{\,\in\,}[0,1]^{M}$ depend on the way the consistency measures are implemented.
Under \texttt{Bayesian}, they are simply those updated through~\eqn{bayes_update}; under \texttt{Infnorm CSN}, we use the rule described in~\remk{consistency}.
The $\mathds{1}\{\text{step 2}\}$ term in~\eqn{robust_margin_bound} indicates this constraint is only included when we are solving the second optimization problem. 
Constraint~\eqn{adaptation_constraint} is an adaptation constraint designed to prevent wildly time-varying controllers; of course, this constraint is excluded at initial time $t{\,=\,}0$.
The pseudocodes for our main methods are presented in Algorithms~\ref{alg:algo23} and~\ref{alg:algo4}.

\begin{algorithm}
\small{
\caption{\small{Stochastic Multimode SLS with Infnorm CSN (No.~2 and~3 in Tab.~\ref{tab:compare_architectures}).}}
\label{alg:algo23}
\begin{algorithmic}[1]
    \STATE{Initialize MJLS, mode $\xi_0$, $\Ccal_i[0]{\,=\,}\{\xi_0\}$, $\overline{\Ccal}_i[0]{\,=\,}\{\xi_0\}$, $\xvect[0]=\hat{\wvect}[0]=\xvect_0$.}
    \FOR{$t=0,\cdots,T_{\text{sim}}-1$}
        \FOR{$i{\,\in\,}\Vcal$}
            \STATE{Default $\boldsymbol{\Phi}_i^{(t)}=\boldsymbol{\Phi}_i^{(t-1)}$.}
            \IF{$t=0$ or $\big($$t\geq 1$ and $\overline{\Ccal}_i[t-1]\neq\overline{\Ccal}_i[t]$$\big)$}
                \STATE{Update $\boldsymbol{\Phi}_i^{(t)}$ using~\eqn{StochAdapt} with $\pvect_i[t]$ and Rmk.~\ref{rmk:consistency}.}
            \ENDIF
        \ENDFOR
        \STATE{Update signals via~\eqn{sf_implement} with $\boldsymbol{\Phi}^{(t)}$.}
        \STATE{Propagate true dynamics~\eqn{plant_dynamics}: $\xvect[t]{\,\to\,}\xvect[t+1]$.}
        \FOR{$i{\,\in\,}\Vcal$}
            \STATE{Update consistent set $\Ccal_i[t]$ via~\eqn{infnorm}.}
            \STATE{Create merged consistent set $\overline{\Ccal}_i[t]$ from~\eqn{merged_consistency_infnorm}.\\
            (\textit{For \texttt{Decentral-Separate} subcase, $\overline{\Ccal}_i[t] = \Ccal_i[t]$.})}
        \ENDFOR
    \ENDFOR
\end{algorithmic}
}
\end{algorithm}
\vspace{-.3cm}
\begin{algorithm}
\small{
\caption{\small{Stochastic Multimode SLS with Bayesian Consistency (No.~4 in Tab.~\ref{tab:compare_architectures}).}}
\label{alg:algo4}
\begin{algorithmic}[1]
    \STATE{Initialize MJLS, mode $\xi_0$, $\pvect_{i,0}$ all zero except $1$ at $\xi_0$, $\xvect[0]=\hat{\wvect}[0]=\xvect_0$.}
    \FOR{$t=0,\cdots,T_{\text{sim}}-1$}
        \FOR{$i{\,\in\,}\Vcal$}
            \STATE{Update $\boldsymbol{\Phi}_i^{(t)}$ using~\eqn{StochAdapt} with $\pvect_i[t]$.}
        \ENDFOR
        \STATE{Update signals via~\eqn{sf_implement} with $\boldsymbol{\Phi}^{(t)}$.}
        \STATE{Propagate true dynamics~\eqn{plant_dynamics}: $\xvect[t]{\,\to\,}\xvect[t+1]$.}
        \FOR{$i{\,\in\,}\Vcal$}
            \STATE{Update $\pvect_i[t+1]$ via~\eqn{bayes_update}.}
        \ENDFOR
    \ENDFOR
\end{algorithmic}
}
\end{algorithm}

\begin{remark}[Mode Identification Tradeoff]\label{rmk:fullrobust}
    We emphasize that mode estimation and consensus do not need to fully identify the true mode of the system in order for our control pipeline to achieve stabilization.
    This is because \texttt{StochAdapt} is designed for robustness: the likelihood probabilities $\pvect_i[t]$ are used as robustness measures that enable simultaneous stabilization across multiple consistent plants.
    This contrasts with many data-driven online control methods that often require a two-step sequential full-identification-then-control procedure.
    One might then wonder about the necessity of having a mode estimation procedure, which leads to a question of determining to what extent we should \textit{trade off} 1) reducing the size of the uncertainty set, and 2) letting the robust controller handle the uncertainty set.
    While we defer a more formal analysis of this tradeoff to future work, we empirically address this concern by implementing an extreme version of the proposed framework, \texttt{FullRobust} in~\tab{compare_architectures}, which uses uniform probabilities $\pvect_{i,m}[t]{\,=\,}1/M$ for all $m{\,\in\,}\Xcal$ and all nodes $i$ for all time.
    The two-step optimization problem~\eqn{StochAdapt} is solved exactly once with these uniform probabilities, and the resulting $\boldsymbol{\Phi}^{(0)}$ response is used for all time.
    Because there is no mode estimation, there is also no need for any consensus.
    Essentially, by removing the mode estimation phase entirely and maintaining the original size of the uncertainty set, \texttt{FullRobust} trains the controller to robustly stabilize all modes with equal importance.
    \sec{simulation} will demonstrate that even though full identification of the true mode is not always necessary, partial identification is still important.    
\end{remark}

\section{Numerical Simulations}\label{sec:simulation}
\begin{figure}
    \centering
    \includegraphics[width=0.95\linewidth]{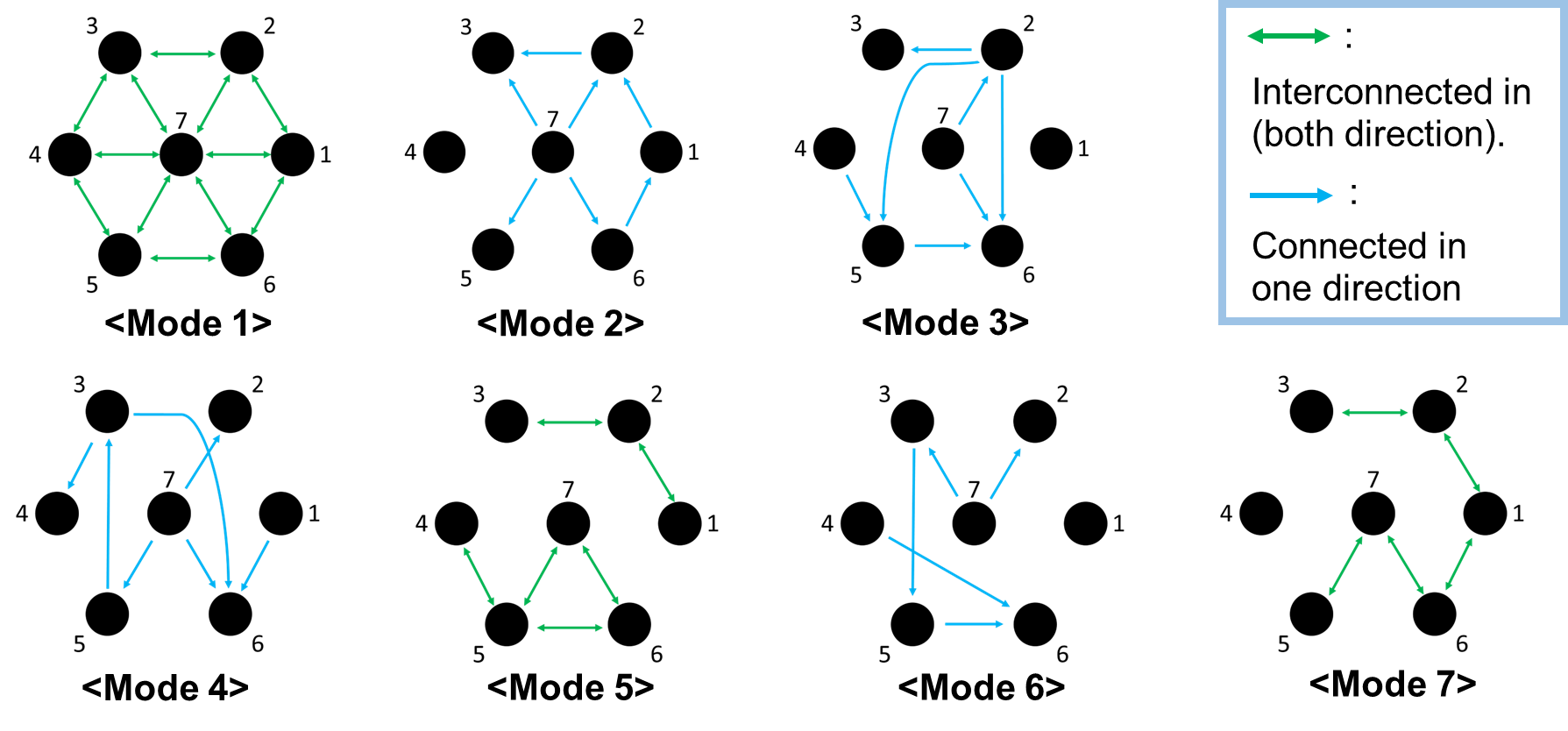}
    \includegraphics[width=0.95\linewidth]{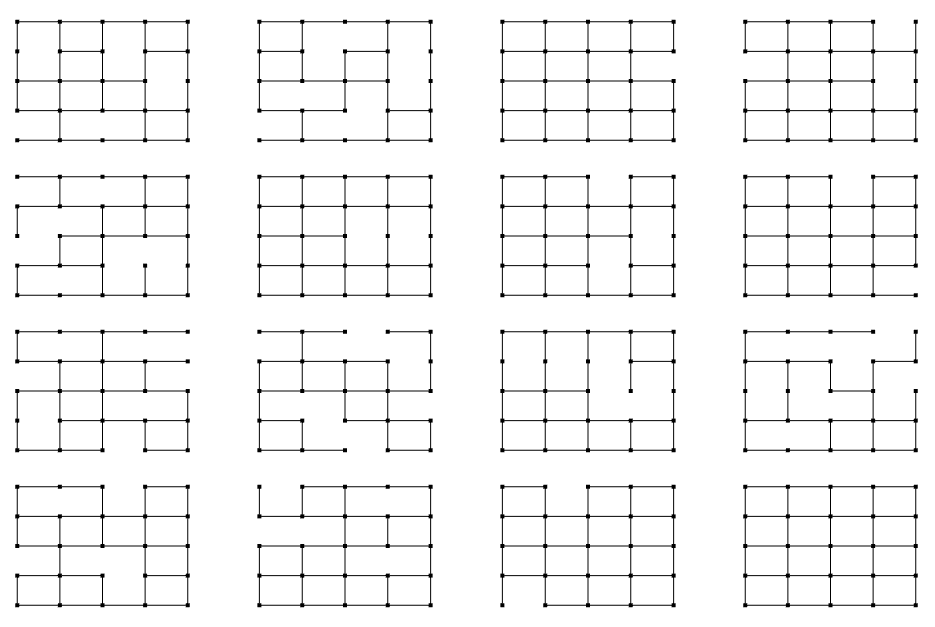}
    \caption{Different modes for [Top] Hexagon, and
    [Bottom] rectangular Grid.
    Some of the interconnections in Grid are asymmetric too, but we omit those details for visual cleanliness.}
    \vspace{-.5cm}
    \label{fig:modes}
\end{figure}

Our methods are simulated on the specific application of a disturbance-rejection in a power grid network.
The system model~\eqn{plant_dynamics_local} is that of the discretized and linearized swing dynamics from~\cite{wang18}:
\begin{align*}
    A_{(ii)}&\triangleq \begin{bmatrix}
   1 &\Delta t \\ -\frac{k_i (m)}{m_i} \Delta t & 1- \frac{d_i}{m_i} \Delta t
    \end{bmatrix},\\
    A_{(ij)}&\triangleq\begin{bmatrix}
        0 & 0 \\ \frac{k_{ij}}{m_i} \Delta t & 0 
    \end{bmatrix}, \hskip.3cm B_{(ii)}\triangleq\begin{bmatrix}
        0 \\ \frac {\Delta t} {m_i}
    \end{bmatrix}
\end{align*}
We choose $\Delta t{\,=\,}0.2$, $d_i{\,=\,}1.5$, $m_i{\,=\,}1$, $k_{ij}{\,=\,}2.5$, and $k_i (m){\, =\,}\sum_{j \in \mathcal {N}_i(m)} k_{ij}$.
Each component of the initial condition is mean-zero Gaussian with standard deviation $\sigma_{x_0}{\,=\,}1$.
The mode switches represent changes in network topology due to line failures (e.g., from extreme weather conditions) and new line installations, and the interarrival times are Geometrically-distributed with parameter $1-e^{-1/5}$.
We do \emph{not} assume that the system interconnection is symmetric: $j{\,\in\,}\mathcal{N}_i(m)$ need not imply $i{\,\in\,}\mathcal{N}_j(m)$.
We perform our comparisons on two different topologies: a (small-scale) Hexagon system and a (large-scale) rectangular Grid system.
The Hexagon system consists of a hexagonal arrangement of $N{\,=\,}7$ nodes and $M{\,=\,}7$ topologies.
The Grid system is a $5\times 5$ rectangular grid of $N{\,=\,}25$ nodes and $M{\,=\,}16$ topologies (see~\fig{modes}).
In both systems, the TPM is a randomly-generated $M{\,\times\,}M$ stochastic matrix.

Our proposed methods (No.~2-4 of~\tab{compare_architectures}) are compared to two baselines (No.~1 and~5 of~\tab{compare_architectures}).
No.~1 is a generalization of our prior approach (i.e.,~\cite{han20l4dc}) which can handle asymmetric adjacency matrices, and uses the support constraints in~\eqn{support_constraints}.
No.~5 is the \texttt{FullRobust} extreme case described in~\remk{fullrobust}.
With \texttt{Infnorm CSN}, we choose uniformly distributed $\norm{\wvect[t]}{\,\leq\,}\overline{w}$ with $\overline{w}{\,=\,}0.3$.
With \texttt{Bayesian}, each noise component is mean-zero Gaussian with standard deviation $\sigma_w{\,=\,}0.2$, and we use thershold values $\varepsilon{\,=\,}0.05$, $R{\,=\,}3$, and $\epsilon{\,=\,}10^{-8}$.
\texttt{StochAdapt} is implemented with $\rho^*{\,=\,}20$, $h{\,=\,}1$, and objective function $g(\boldsymbol{\Phi}^{(t)}){\,=\,}\sum_{\tau=1}^T \sum_{i=1}^N ( \lVert\Phi_{x,i}^{(t)}[\tau]\rVert + 0.1\lVert\Phi_{u,i}^{(t)}[\tau]\rVert)$, where $\norm{\cdot}$ is the Frobenius norm.
Our performance metrics for the controllers are the time-normalized $l^2$~norm of the state and control trajectories, respectively $X{\,\triangleq\,}(1/T_{\text{sim}})\sum_{t=1}^{T_\text{sim}} \norm{\xvect[t]}^2$ and $U{\,\triangleq\,}(1/T_{\text{sim}})\sum_{t=0}^{T_\text{sim}-1} \norm{\uvect[t]}^2$. Our performance metric for information consensus and the consistency measures is the proportion of having a correct estimate of the mode, expressed as $\hat{\xi}{\,\triangleq\,}(1/T_{\text{sim}})\sum_{t=0}^{T_\text{sim}-1}\mathds{1}\{\xi_t{\,=\,}\hat{\xi}_t\}$, where $\hat{\xi}_t$ represents the result of consensus among all $N$ subsystems.

\begin{figure}
    \centering
    \includegraphics[width=\linewidth]{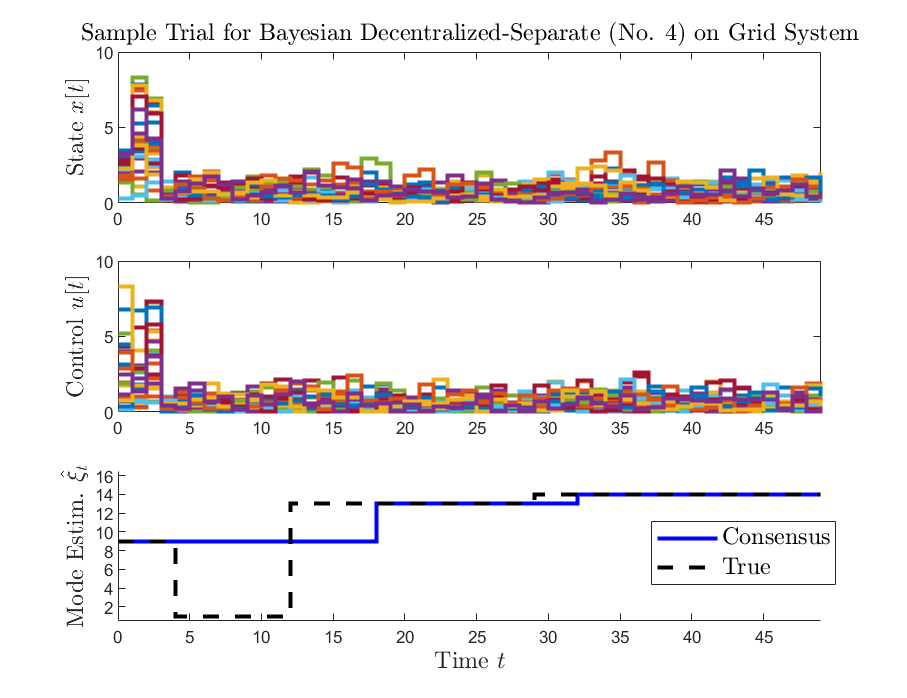}
    \vspace{-.3cm}\includegraphics[width=\linewidth]{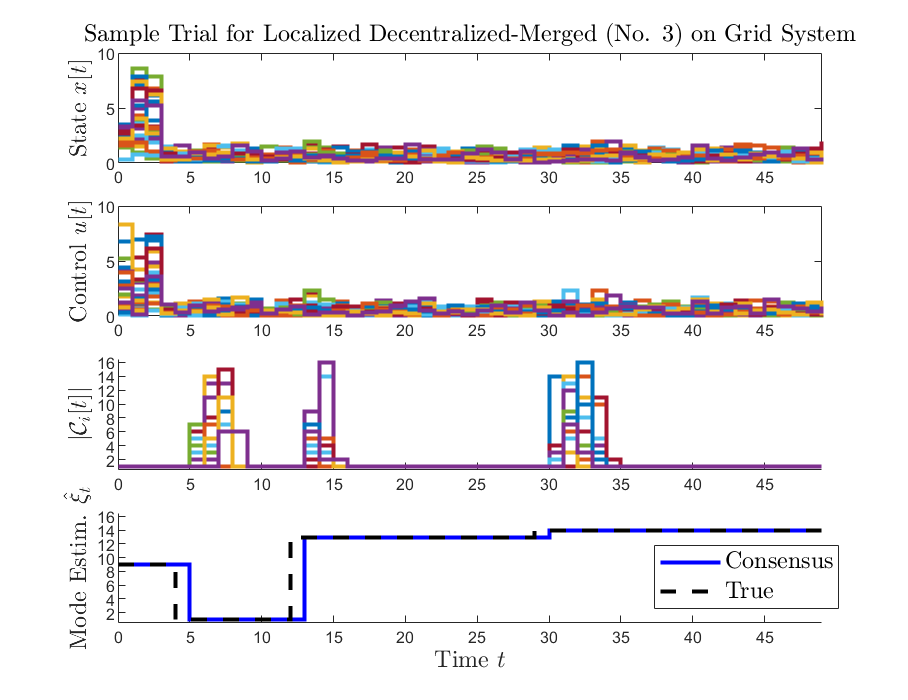}
    \caption{Sample trajectories for algorithms No.~3 and 4 in Tab.~\ref{tab:compare_architectures} implemented on the Grid system.}
    \vspace{-.5cm}
    \label{fig:sample_trials}
\end{figure}

Our results are summarized in~\tab{results}; some sample trajectories are plotted in~\fig{sample_trials}.
For each case, we performed $20$ Monte-Carlo trials over a simulation duration of $T_{\text{sim}}{\,=\,}50$ timesteps.
All simulations were run on an Intel i9-13900K PC with 128GB RAM and parallel processing on 6 cores.
We emphasize that parallel processing was used only to run multiple trials at once, and that a less powerful system (e.g., an Intel i7-10510U PC with 16GB RAM) can easily run a single trial.
Deployment on a real distributed system is expected to speed up computation further, since each subsystem is a separate PC that runs independently of the other subsystems outside of occasional communication.

Both systems loosely follow the same trends.
First, as expected, No.~1 (with \texttt{Infnorm CSN} and \texttt{Centralized} consensus) has the best coherence among the subsystems due to maintaining only a single global estimate of the true mode.
Second, also expected, No.~4 has the worst coherence since it implements \texttt{Decentral-Separate} consensus and because \texttt{Bayesian} consistency assigns nonzero weights across all modes $\Xcal$ due to the unbounded nature of Gaussian-distributed $\wvect[t]$.
Third, both No.~2 and No.~3 (the decentralized consensus algorithms) expend slightly less control effort compared to No.~1, and the inclusion of consensus improves the disturbance-rejection performance $X$ slightly.
%
We note there is better disturbance rejection $X$ for Hexagon, a smaller-scale system where taking the immediate neighbors of each subsystem is close to covering the entire network, but it is also more difficult to control than Grid, as we see there is little difference in $U$ despite having less nodes.
Controller architectures with \texttt{Infnorm CSN} ran more quickly than those with \texttt{Bayesian} consistency implemented because we skip iterations without recomputing $\boldsymbol{\Phi}_i^{(t-1)}$ if the consistent set $\Ccal_i[t]{\,=\,}\Ccal_i[t{\,-\,}1]$ is unchanged.

Finally, for No.~5 (\texttt{FullRobust}) applied to the Hexagon system perturbed by norm-bounded noise $\wvect[t]$, we observe complete destabilization with $X = 83.542(\pm90.398)$, $U=1.213\times 10^{-6}(\pm1.499\times 10^{-6})$.
Applying No.~5 to Gaussian-distributed noise $\wvect[t]$ and the Grid system yields similar results.
This suggests that simultaneously stabilizing across too many modes for the entire time duration causes the controller to ``give up'' ($u{\,\simeq\,}0$), and indeed confirms the necessity of the \emph{partial} mode estimation, as seen by~\tab{results}.

\bgroup
\def\arraystretch{1.2}
\begin{table}
    \begin{adjustbox}{width=0.95\columnwidth,center}
      \begin{tabular}{| c | c | c | c | c |}
        \hline
         No. & 1 & 2 & 3 & 4\\ \hline
         $X$ & $6.500(\pm0.85)$ & $6.627(\pm0.82)$ & $6.533(\pm0.90)$ & $8.232(\pm1.73)$\\ \hline
         $U$ & $14.460(\pm1.58)$ & $14.257(\pm1.61)$ & $14.414(\pm1.63)$ & $14.906(\pm2.25)$\\ \hline
         $\hat{\xi}$ & $0.998(\pm0.02)$ & $0.930(\pm0.02)$ & $0.937(\pm0.02)$ & $0.784(\pm0.08)$ \\ \hline\hline
      \end{tabular}
    \end{adjustbox}

    \begin{adjustbox}{width=0.95\columnwidth,center}
      \begin{tabular}{| c | c | c | c | c |}
        \hline
         No. & 1 & 2 & 3 & 4 \\ \hline
         $X$ & $24.317(\pm 0.76)$ & $24.488(\pm 0.75)$ & $24.338(\pm 0.73)$ & $26.529(\pm 1.10)$\\ \hline
         $U$ & $16.168(\pm 0.65)$ & $16.079(\pm 0.66)$ & $16.172(\pm 0.68)$ & $17.372(\pm 0.83)$\\ \hline
         $\hat{\xi}$ & $0.996(\pm0.05)$ & $0.911(\pm0.05)$ & $0.913(\pm0.05)$ & $0.676(\pm0.10)$ \\ \hline
      \end{tabular}
    \end{adjustbox}
    \caption{Monte-Carlo averaged performance metrics for each system. [Top] Hexagon. [Bottom] Grid.
    No.~5 was unstable for both systems and noise profiles, and is omitted here.
    }
    \label{tab:results}
\end{table}
\egroup

\section{Conclusion}\label{sec:conclusion}
We proposed various controller architectures for robustly stabilizing large-scale MJLS under external disturbances and topological changes while adaptively estimating the unknown mode.
Our main pipeline is a two-step convex optimization procedure~\eqn{StochAdapt} which simultaneously stabilizes multiple plants that are consistent with the observable data.
A key component of our framework is the computation of a likelihood distribution across the system's possible modes, which were used as priority weights in~\eqn{StochAdapt} and updated via Bayesian inference.
Scalability to larger networks was achieved by localized controller implementation and information consensus.
Future work includes deployment on real (and larger) distributed systems, merged consensus for \texttt{Bayesian} consistency,
and some theoretical analyses on~\eqn{StochAdapt}, such as mode-estimation convergence properties
and formalization via dynamic programming.

\bibliographystyle{IEEEtran}
\bibliography{%
bibi%
}

\begin{thebibliography}{10}
\providecommand{\url}[1]{#1}
\csname url@samestyle\endcsname
\providecommand{\newblock}{\relax}
\providecommand{\bibinfo}[2]{#2}
\providecommand{\BIBentrySTDinterwordspacing}{\spaceskip=0pt\relax}
\providecommand{\BIBentryALTinterwordstretchfactor}{4}
\providecommand{\BIBentryALTinterwordspacing}{\spaceskip=\fontdimen2\font plus
\BIBentryALTinterwordstretchfactor\fontdimen3\font minus
  \fontdimen4\font\relax}
\providecommand{\BIBforeignlanguage}[2]{{%
\expandafter\ifx\csname l@#1\endcsname\relax
\typeout{** WARNING: IEEEtran.bst: No hyphenation pattern has been}%
\typeout{** loaded for the language `#1'. Using the pattern for}%
\typeout{** the default language instead.}%
\else
\language=\csname l@#1\endcsname
\fi
#2}}
\providecommand{\BIBdecl}{\relax}
\BIBdecl

\bibitem{aberkane2011}
S.~Aberkane, ``Stochastic stabilization of a class of nonhomogeneous
  {M}arkovian jump linear systems,'' \emph{Systems \& Control Letters},
  vol.~60, no.~3, pp. 156--160, 2011.

\bibitem{matei2008}
I.~Matei, N.~C. Martins, and J.~S. Baras, ``Optimal linear quadratic regulator
  for {M}arkovian jump linear systems, in the presence of one time-step delayed
  mode observations,'' \emph{IFAC Proceedings Volumes}, vol.~41, no.~2, pp.
  8056--8061, 2008, 17th IFAC World Congress.

\bibitem{deSouza2006}
C.~de~Souza, ``Robust stability and stabilization of uncertain discrete-time
  {M}arkovian jump linear systems,'' \emph{IEEE Transactions on Automatic
  Control}, vol.~51, no.~5, pp. 836--841, 2006.

\bibitem{anderson19}
J.~Anderson, J.~C. Doyle, S.~H. Low, and N.~Matni, ``System level synthesis,''
  \emph{Annual Reviews in Control}, vol.~47, pp. 364--393, 2019.

\bibitem{han20l4dc}
S.~Han, ``Localized learning of robust controllers for networked systems with
  dynamic topology,'' in \emph{Pro. 2nd Conf. Learning Dynamics Control}, ser.
  Proceedings of Machine Learning Research, vol. 120.\hskip 1em plus 0.5em
  minus 0.4em\relax PMLR, Jun 2020, pp. 687--696.

\bibitem{costa_book}
O.~Costa, M.~Fragoso, and R.~Marques, \emph{Discrete-Time Markov Jump Linear
  Systems}, ser. Applied probability.\hskip 1em plus 0.5em minus 0.4em\relax
  Springer, 2005.

\bibitem{wang19}
Y.-S. Wang, N.~Matni, and J.~C. Doyle, ``A system-level approach to controller
  synthesis,'' \emph{IEEE Transactions on Automatic Control}, vol.~64, no.~10,
  pp. 4079--4093, 2019.

\bibitem{wang18}
Y.-S. Wang, N.~Matni, and J.~Doyle, ``Separable and localized system level
  synthesis for large-scale systems,'' \emph{IEEE Transactions on Automatic
  Control}, vol.~63, no.~12, pp. 4234--4249, Dec 2018.

\bibitem{olfati04}
R.~{Olfati-Saber} and R.~M. {Murray}, ``Consensus problems in networks of
  agents with switching topology and time-delays,'' \emph{IEEE Transactions on
  Automatic Control}, vol.~49, no.~9, pp. 1520--1533, Sep 2004.

\bibitem{xiao05}
L.~{Xiao}, S.~{Boyd}, and S.~{Lall}, ``A scheme for robust distributed sensor
  fusion based on average consensus,'' in \emph{IPSN 2005. Fourth International
  Symposium on Information Processing in Sensor Networks, 2005.}, Apr 2005, pp.
  63--70.

\bibitem{han23auto}
S.~Han, S.-J. Chung, and J.~C. Doyle, ``{Predictive control of linear
  discrete-time Markovian jump systems by learning recurrent patterns},''
  \emph{Automatica}, vol. 156, p. 111197, Oct 2023.

\bibitem{doucet2008}
\BIBentryALTinterwordspacing
A.~Doucet and A.~M. Johansen, ``A tutorial on particle filtering and smoothing
  : {F}ifteen years later,'' 2008. [Online]. Available:
  \url{https://www.stats.ox.ac.uk/~doucet/doucet_johansen_tutorialPF2011.pdf}
\BIBentrySTDinterwordspacing

\bibitem{blondel92}
V.~Blondel, G.~Campion, and M.~Gevers, ``A sufficient condition for
  simultaneous stabilization,'' \emph{IEEE Transactions on Automatic Control},
  vol.~38, no.~8, pp. 1264--1266, 1993.

\bibitem{cao99}
Y.-Y. Cao, Y.-X. Sun, and J.~Lam, ``Simultaneous stabilization via static
  output feedback and state feedback,'' \emph{IEEE Transactions on Automatic
  Control}, vol.~44, no.~6, pp. 1277--1282, 1999.

\bibitem{blondel94_book}
V.~Blondel, \emph{Simultaneous Stabilization of Linear Systems}, ser. Lecture
  Notes in Control and Information Sciences.\hskip 1em plus 0.5em minus
  0.4em\relax Springer-Verlag, 1994.

\bibitem{fonte2001}
C.~Fonte, M.~Zasadzinski, C.~Bernier-Kazantsev, and M.~Darouach, ``On the
  simultaneous stabilization of three or more plants,'' \emph{IEEE Transactions
  on Automatic Control}, vol.~46, no.~7, pp. 1101--1107, 2001.

\bibitem{costa95}
O.~Costa and M.~Fragoso, ``Discrete-time {LQ}-optimal control problems for
  infinite {M}arkov jump parameter systems,'' \emph{IEEE Transactions on
  Automatic Control}, vol.~40, no.~12, pp. 2076--2088, 1995.

\bibitem{costa98}
O.~Costa and R.~Marques, ``Mixed {$\mathcal{H}_2$}/{$\mathcal{H}_{\infty}$}
  control of discrete-time {M}arkovian jump linear systems,'' \emph{IEEE
  Transactions on Automatic Control}, vol.~43, no.~1, pp. 95--100, 1998.

\end{thebibliography}

\end{document}